# Naturalistic Yielding Behavior of Drivers at an Unsignalized Intersection based on Survival Analysis

Delgermaa Gankhuyag, Cristina Olaverri -Monreal, Senior *Member, IEEE*

*Abstract—* In recent years, autonomous vehicles have become increasingly popular, leading to extensive research on their safe and efficient operation. Understanding road yielding behavior is crucial for incorporating the appropriate driving behavior into algorithms. This paper focuses on investigating drivers' yielding behavior at unsignalized intersections. We quantified and modelled the speed reduction time for vulnerable road users at a zebra crossing using parametric survival analysis. We then evaluated the impact of speed reduction time in two different interaction scenarios, compared to the baseline condition of no interaction through an accelerated failure time regression model with the log-logistic distribution. The results demonstrate the unique characteristics of each yielding behavior scenario, emphasizing the need to account for these variations in the modelling process of autonomous vehicles.

## I. INTRODUCTION

Intersections are an essential part of road infrastructure. Signalized intersections, also known as controlled intersections, have traffic signals that regulate the flow of traffic. In contrast, unsignalized intersections lack traffic signals and rely on road users' judgement to navigate safely.

Drivers behavior at intersections varies depending on the type of intersection they encounter. At signalized intersections, drivers tend to follow traffic rules and regulations. However, at unsignalized intersections, drivers exhibit more naturalistic behavior and may not always follow strict rules. For example, they may yield to pedestrians or other vehicles even when vehicles have the right of way.

Controlled environments, where unpredictable behavior is minimized, make it easier for self-driving cars to interact with other road users who adhere to traffic rules and regulations.

On the other side, scenarios that include unsignalized intersections can be particularly challenging for autonomous vehicles, as they have to rely on sensors and algorithms to navigate safely and avoid collisions. This can be especially dangerous for vulnerable road users, such as pedestrians and cyclists. Autonomous vehicles (AVs) have gained significant popularity worldwide in recent years, and extensive research is being conducted to evaluate the social acceptance of this groundbreaking technology. While human drivers use nonverbal gestures, such as hand and head movements, to interact safely with other road users [1], the interaction and communication between road users and AVs are determined solely by their movement, distance and time-based factors. This is particularly challenging at unsignalized intersections, where road users and driverless cars might face difficulties interacting with each other. Unlike human drivers, AVs follow strict, rule-based algorithms and might not let vulnerable road users cross the road at unsignalized intersections.

It is crucial to understand these differences and incorporate the corresponding driving behavior into the design and programming of autonomous vehicles to ensure their safe and efficient operation, improve trust in them, and promote greater social acceptance of this innovative technology.

Therefore, we contribute to the existing knowledge by analyzing the yielding behavior of human drivers in vehicles at unsignalized intersections to better understand the interaction challenges and identify ways to improve them.

The remainder of this paper is structured as follows: Section II provides an overview of the relevant literature. Section III outlines our methodology, including the description of the statistical method employed. In Section IV, we present the results of our analysis. Finally, in Section V, we conclude the work and outline future research.

## II. RELATED WORK

The interaction between pedestrians and driverless vehicles has been investigated in previous work through Wizard of Oz experiments [2] and virtual reality experiments [3] at intersections and zebra crossing scenarios. Results showed that yielding behavior and size affected the pedestrians decision to cross.

Further research focused on using real driverless vehicles for field tests in intersections and shared spaces, adhering to external communication messages and focusing on the promotion of trust for the public's acceptance of the new self-driving car technology. [4, 5, 6, 7].

Simulation studies have complemented the work at crosswalks using drivers' speed reduction time as a parameter to determine yielding behavior at bicycle crossroads using a survival model [8]. This particular work focuses on bicycles on a simulated environment. Other parameters that have been identified to analyze interactions and yielding behavior are the time to collision (TTC) and vehicle deceleration in unmarked and marked crosswalks [9, 10]. Results from previous research at unsignalized intersections have shown that pedestrians start crossing when the time gap between them and vehicles approaching the zebra crossing is 5 seconds or more [11].

While there have been numerous studies examining pedestrian-vehicle interactions, there is a limited number of papers focusing on cyclist-vehicle interactions at

Chair ITS-Sustainable Transport Logistic 4.0, Johannes Kepler University Linz. delgermaa.gankhuyag@jku.at, cristina.olaverri-monreal@jku.at

intersections. The safe interaction between cyclists and vehicles is greatly influenced by factors such as the initial speed of both cyclists and drivers, the time difference in arrival at an intersection, and visual cues exhibited by cyclists, such as pedaling and maintaining eye contact with drivers [12]. However, cyclists and pedestrians generally interact smoothly with AVs at intersections unless the vehicles exhibit aggressive or unexpected behaviors, such as not yielding at crosswalks or stopping too close to vulnerable road users [2].

Not all the studies described in this section focused on a specific intersection design, as some aimed to observe the natural yielding behavior of drivers through continuous recording. Additionally, the analysis in some previous work did not consider the natural deceleration behavior when there was no interaction between drivers and VRUs, sometimes resulting in the absence of a baseline condition.

We investigate drivers yielding behaviour toward pedestrians and cyclists at an unsignalized intersection with a zebra crossing. We contribute to field of research by defining our own model distribution based on the behaviour of the data that we analysed from a bird's-eye-view naturalistic driving dataset.

## III. METHODOLOGY

To analyze the yielding behavior of human drivers in vehicles at unsignalized intersections, we relied on the bird's-eye view driving scenarios in Aachen, Germany, that are available in the *inD* dataset [13]. We selected a unsignalized four-way intersection near the city center, known as Frankenburg, due to its high volume of vulnerable road users. The total recording duration was approximately four hours, with a speed limit of 15 meters per second. The intersection features a zebra crossing, where most of the interactions with vulnerable road users take place. To detect the naturalistic yielding behavior at the zebra crossing at the intersection, the following three scenarios were chosen:

**Baseline scenario** – Drivers do not interact with any road users at the intersection. To observe the naturalistic yielding behavior of drivers towards vulnerable road users, the "No interaction" scenario is considered as a baseline behavior due to the location of the zebra crossing right next to the unsignalized four-way intersection conflict zone. As drivers approach the conflict zone - the middle section of the intersection - they tend to naturally brake (see Fig. 1).

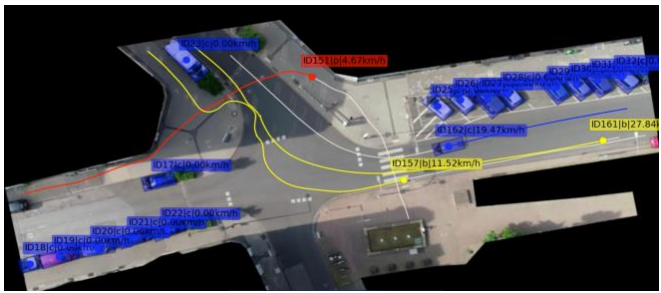

Figure 1. Baseline scenario where a car drives through an intersection without interacting with other road users.

**Drivers and pedestrians interaction scenario** – At the zebra crossing, drivers are expected to yield to pedestrians who wish to cross the road. In order to ensure consistency and avoid introducing any bias, the analysis performed in this work only includes instances where cars approach the crossing from the right-hand side. This enables a more homogeneous dataset for each scenario and facilitates accurate analysis. The information visualized in Fig. 2 provides a better understanding of the interaction between the car and the pedestrian.

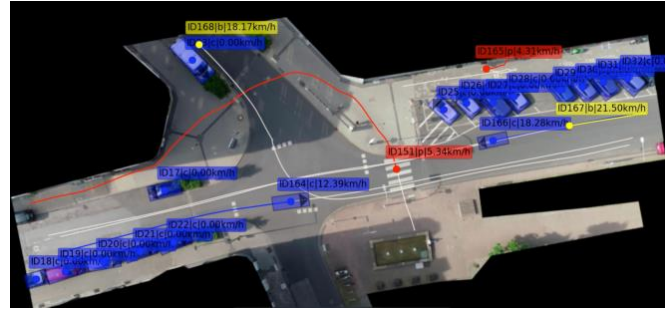

Figure 2. A car (with id 166) can be seen yielding to a pedestrian at the unsignalized intersection. The red dot in the picture represents the pedestrian, while the blue boxes denote the cars. The solid red, blue, and yellow lines indicate the historical trajectory of the road users, while the solid grey lines depict the future trajectories.

**Driver and cyclists interaction scenario** – Although there is no designated cycle path at the intersection, cyclists often use the zebra crossing to cross the road. Therefore, in our analysis, we take into account the yielding behavior of drivers towards both pedestrians and cyclists. Fig. 3 depicts the scenario.

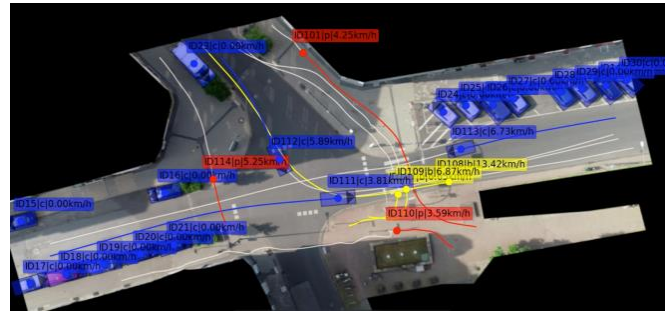

Figure 3. A scenario where a car (with id 113) is yielding to cyclists at an unsignalized intersection. In the picture, two yellow dots indicate the presence of cyclists who are crossing the road using the zebra crossing. The solid red, blue, and yellow lines in the picture show the historical trajectory of the road users, while the solid grey lines depict their future trajectories.

To quantify yielding behavior in a naturalistic setting, we have used speed reduction time. Based on this approach, we present the following hypothesis for the scope of this paper.

### A. Hypotheses

H0: There is no significant difference in the average speed reduction time (SRT) of vehicles when interacting with vulnerable road users, including cyclists and pedestrians, at an unsignalized intersection's zebra crossing.

H1: There is a difference in the average speed reduction time (SRT) of at least one vehicle when interacting with different road users at a zebra crossing in an unsignalized intersection. Specifically, the SRT varies among different types of road users, such as cyclists and pedestrians.

### B. Sample size

After classifying the scenarios in the dataset, we selected 211 instances where drivers did not interact with any road users, 63 scenarios where drivers yielded to pedestrians at the zebra crossing, and 9 cases where drivers yielded to cyclists at the zebra crossing. In order to ensure the validity of our conclusions, we refrained from increasing the sample size by mixing data from different locations or using simulated data, which could result in biased or unrealistic results. Instead, we focused on a single location and utilized the available data to establish a homogeneous dataset for our analysis.

### C. Modelling and assumptions

Survival analysis is a statistical method used to analyze data in which the outcome in interest is **"time-to-event"**. Survival time refers to the length of time between starting point and an endpoint, where the endpoint is typically a specific event of interest. The time can be in any metric such as seconds, days, weeks, etc. An event is any occurrence depending on the research question; for instance, it could be taking over maneuver, braking, turning, or stopping in the automotive field. In survival analysis, the goal is to estimate the probability of an event occurring over time and identify factors that may be associated with the occurrence of the event [14].

Let **T** denote the survival time. In our application, **T** is SRT, which is braking time from maximum to minimum speed to VRU before a zebra crossing. The survival function $S_T(t)$ is the probability that the event, *drivers reaching to the minimum speed before the zebra crossing*, occurs later than some time t and is defined as:

$$S_T(t) = P(\mathbf{T} > t) \quad (1)$$

where t is any non-negative value in seconds.

The accelerated failure time (AFT), parametric survival model, explains a linear relationship between the logarithm of the survival time and the covariates. For a given SRT and a vector of covariates $\mathbf{X} \in \mathbb{R}^p$ with corresponding parameters $\boldsymbol{\beta} \in \mathbb{R}^p$, the AFT model can be formulated on the log-scale as

$$\mathbf{Y} = \beta_0 + \boldsymbol{\beta}'\mathbf{X} + \sigma\varepsilon \quad (2)$$

where $\mathbf{Y} = \log(t = SRT)$, $\varepsilon$ is a random error term assumed to follow some distribution, and $\beta_0$ is the intercept. $\sigma$ follows a standardized distribution.

In the AFT model, the survival time is modelled as a function of covariates by transforming the baseline survival time. The survival function can be expressed as

$$S(t|x)=[S_0(g(x)t)] \quad (3)$$

where $S(t|x)$ is the survival function for a driver with covariates value x at time t, $S_0(t)$ is the baseline survival function, and $g(x)$ is the function of covariates. When $g(x) > 1$, time is accelerating, and $g(x) < 1$, time is decelerating. If $g(x) = \exp(\mathbf{X}\boldsymbol{\beta})$, which results in a linear regression model for $\ln(t)$. Therefore, the survival function is $S(t|x)=[S_0(\exp(\mathbf{X}\boldsymbol{\beta})t)]$.

The main assumption of the AFT model has a multiplicative effect of covariates on survival time, meaning each scenario's SRT differs by $t_1 = \exp(\beta_j)t_2$ where $\beta_j$ is the jth coefficient of a variable. If we denote the term $\exp(\beta_j)$ by $\gamma$, this term is called the acceleration factor. The acceleration factor is a key measure of the relationship obtained in an AFT model when comparing one group to another with respect to time.

The survival function changes depending on the distribution of survival time in the modelling. The distribution fitting will be discussed in section IV.

### D. Response and covariate variables

All model variables are estimated based on the methodology described in a paper [8].

#### 1) Response variable:
**SRT:** speed reduction time, the elapsed time to transition from the initial maximum speed to the minimum speed.

#### 2) Covariate variables:
**$V_i$:** initial speed, which is the speed identified when the driver starts to decrease the speed in response to the cyclists or pedestrians who are crossing.
**$LV_i$:** initial speed distance, which is the distance from the zebra crossing where $V_i$ is measured.
**$V_m$:** minimum speed, which is the minimum speed during the braking maneuver before the zebra crossing in response to the cyclists or pedestrians crossing.
**$LV_m$:** minimum speed distance, the distance from the zebra crossing where $V_m$ is measured.
**dav:** average deceleration, which is the average deceleration adopted by the driver during the entire braking maneuver as described in equation (4).

$$dav = \frac{(V_i^2 - V_m^2)}{2*(LV_i - LV_m)} \quad (4)$$

**Mtype:** Maneuver type as a categorical variable with three different levels: drivers going straight, turning right, and turning left.

**IType:** Drivers' interaction type as a categorical variable with three levels: no interaction, interaction with pedestrians and cyclists at the intersection.

### E. Variable selection

Based on the backward and forward stepwise variable selection method, the best simple model includes all linear terms of covariates without confounding effects between them. The distribution fitting procedure for the models in variable selection is described in section IV.

The best linear model for the accelerated log-logistic failure time model can be expressed using the following formula:

$$\log(SRT)_j = \beta_0 + \beta_1(V_m)_j + \beta_2(LV_i)_j + \beta_3(LV_m)_j + \\ \beta_4(dav)_j + \beta_5(MType)_j + \beta_6(InterType)_j + \sigma\varepsilon_j \quad (5)$$

where the subscript **j** = {1,……, n} refers to the total number of drivers and $\beta_0, \beta_1, \beta_2, \beta_3, \beta_4, \beta_5, \beta_6$ are the coefficients for the covariates $V_m$, $LV_i$, $LV_m$, dav, Mtype, InterType. $\sigma$ follows a standard logistic distribution, and $\varepsilon$ represents the random error term.

The accelerated lognormal failure time model assumes a simpler relationship between the covariates and the SRT as it has one fewer variable compared to the distributed model with the log-logistic distribution. Hence, the best linear model of lognormal distribution can be represented by the following formula:

$$\log(SRT)_j = \beta_0 + \beta_1(V_m)_j + \beta_2(LV_i)_j + \beta_3(dav)_j + \beta_4(MType)_j + \beta_5(InterType)_j + \sigma\varepsilon_j \quad (6)$$

where the subscript $j = \{1,\ldots\ldots,n\}$ represents the total number of drivers and $\beta_0, \beta_1, \beta_2, \beta_3, \beta_4, \beta_5$ the coefficients for the covariates $V_m$, $LV_i$, dav, Mtype, InterType. $\sigma$ follows a standard normal distribution, and $\varepsilon$ represents the random error term.

According to the multicollinearity analysis in Section IV, the variable $V_i$ representing the initial speed of drivers was not included in the variable selection process to mitigate potential issues arising from multicollinearity, which can lead to unreliable regression coefficients and reduced model interpretability.

## IV. RESULTS

### A. Descriptive analysis

Data from a total of 283 drivers were analyzed to observe natural yielding behaviors in three scenarios as defined in section III: no interaction, encountering pedestrians, and approaching cyclists at a zebra crossing. After obtaining a graphical visualization of the data, Fig. 4, reveals a noticeable contrast in the naturalistic yielding behavior of drivers towards pedestrians and cyclists compared to the scenario where there is no interaction with vulnerable road users.

Although no physical interaction occurs, we observe that as cars approach the conflict zone, they tend to brake quickly and abruptly, while the yielding process for pedestrians and cyclists is smoother and takes a longer time.

Table 1 shows the average values of all numerical variables. When there is no interaction with cars at the zebra crossing, drivers start braking from a distance of 39.81 meters away from the crossing, travelling at a speed of 8.53 meters per second. Their speed decreases until they reach 5.29 meters per second when they are 8.79 meters away from the zebra crossing. In contrast to the no-interaction scenario, when drivers encounter pedestrians at the zebra crossing, they begin to decelerate from approximately 42.97 meters away from the crossing, travelling at an average speed of 7.96 meters per second. The deceleration continues until they reach a distance of 7.61 meters, at which point the cars' speed reaches a minimum average of approximately 1.29 meters per second.

Notably, the yielding behavior towards cyclists in the *interacting with cyclist scenario* exhibits similarities in terms of average values across all variables, with the exception of average deceleration, ***dav***, which is approximately half compared to both the *no-interaction* baseline and *interacting with pedestrians* scenarios. Furthermore, in the *no-interaction* scenario, the speed reduction time is 4.59 seconds, which is less than both *the interacting with pedestrians* (7.15 seconds) and *interacting with cyclists* (8.00 seconds) scenarios. These findings highlight the differences in yielding behaviors and speed reduction times between various road user interactions, emphasizing the unique characteristics of each scenario.

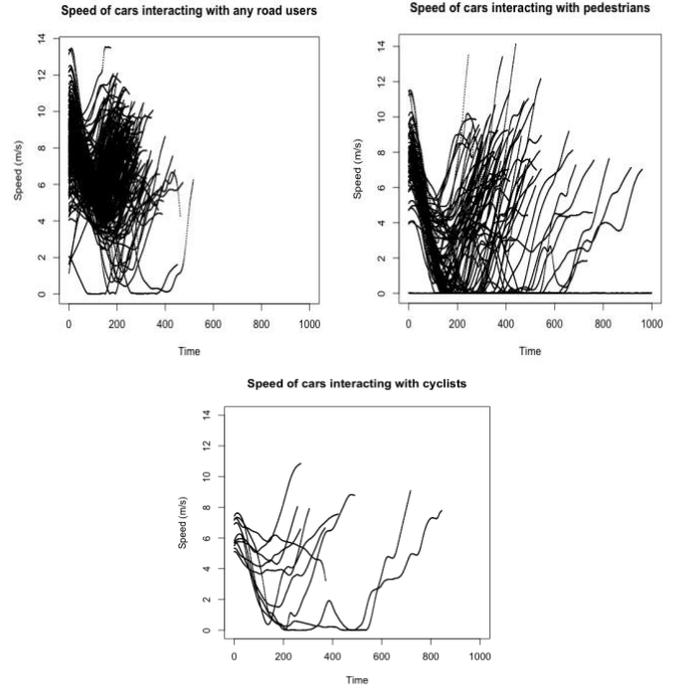

Figure 4. Visualization of the speed of cars at an unsignalized intersection under three different scenarios: no interaction and interaction with vulnerable road users. The top left graph represents the no-interaction scenario, while the top right and bottom plots show the interaction between cars and pedestrians and cars and cyclists, respectively. The time interval for all plots is one second, divided into 25 frames. The dataset includes all driving directions, such as straight, left, and right turns. To facilitate visual comparison, all plots have the same time and speed range.

### B. Multicollinearity analysis

To ensure the reliability and interpretability of the regression model, it is crucial to assess multicollinearity among the covariates. Multicollinearity can introduce challenges such as unstable regression coefficients, diminished predictive accuracy, and reduced interpretability of the model. In Fig. 5, a correlation matrix displays the correlation values between the covariates using a heatmap graph.

TABLE 1. Scenario-based average values of variables

|  | $V_i$ (m/s) | $LV_i$ (m) | $V_m$ (m/s) | $LV_m$ (m) | dav (m/s$^2$) | SRT (s) |
|---|---|---|---|---|---|---|
| No interaction | 8.53 | 39.81 | 5.26 | 8.79 | 0.75 | 4.59 |
| Interacting with pedestrians | 7.96 | 42.97 | 1.29 | 7.61 | 0.88 | 7.15 |
| Interacting with cyclists | 6.39 | 42.43 | 2.19 | 8.17 | 0.48 | 8.00 |

NOTE: Variable units: m/s = meters per second, m = meters, m/s$^2$ = meters per second$^2$, s = seconds

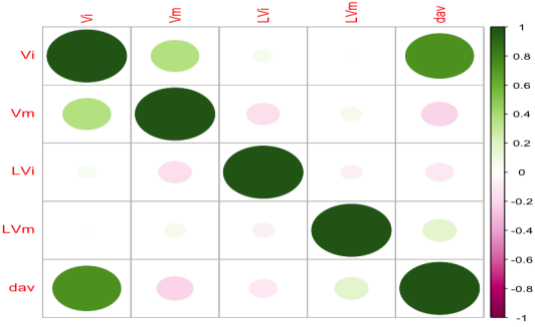

Figure 5. Correlation matrix of the independent variables showing the multicollinearity and correlations that may exist among the covariates.

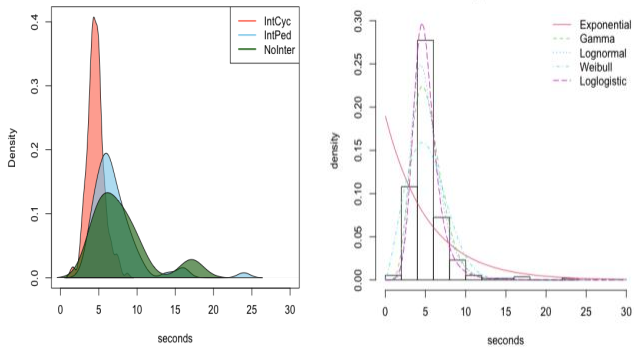

Figure 6. The left graph displays the density plots of speed reduction time for selected scenarios combined in a single graph. In the caption section, **IntCyc** refers to drivers interacting with cyclists; **IntPed** refers to drivers interacting with pedestrians at the zebra crossing; **NoInter** refers to the no-interaction scenario. The right graph shows the empirical and possible theoretical densities fitted to the speed reduction time variable. Note that the range of the "y" axis differs between the two graphs due to the utilization of different visualization packages. Consequently, the estimations may vary slightly from each other. Nevertheless, it is important to emphasize that the same dataset was used for plotting both graphs, ensuring consistency in the visual representation of the data.

Based on the correlation analysis, a strong positive correlation of 73% is observed between the initial speed variable ($V_i$,) and the average deceleration variable (**dav**). Considering the influence of these variables on the outcome variable of interest, which is SRT, it is necessary to exclude the initial speed variable from the modelling process. This decision is made to prevent the presence of unreliable coefficients in the model, as the strong correlation between initial speed and average deceleration can introduce multicollinearity issues and hinder the interpretability of the results.

### C. Distribution fitting

In the parametric survival model, the SRT is assumed to conform to a specific family of distributions [15]. Therefore, we need to define the distribution for the AFT model. Fig. 6 shows the density plots of the response variable for each scenario, combined into a single graph, and the empirical and theoretical densities. Upon observing the left plot, it becomes evident that the yielding behaviors in each scenario originate from distinct distributions rather than a single distribution.

To test and statistically validate the hypothesis, we considered a range of potential general distributions for the outcome variable of interest. By defining these distributions, we aimed to determine whether they are distinct representations of the yielding behaviors or if they can be accounted for through the modelling process.

The speed reduction time variable shows a right-skewed distribution, meaning that the majority of the data is concentrated on the left side of the distribution, and the tail extends towards the higher values. Therefore, to capture the shape of the variable, we fitted five possible theoretical distributions: Exponential, Weibull, Lognormal, Gamma, and Log-logistic. Among these distributions, the Log-logistic distribution fits very well because it covers the entire range of the empirical distribution of speed reduction time. To assess the quality of the fitted distribution, we calculated the Akaike and Bayesian Information Criteria, which provide measures of the goodness of fit (Table 2).

TABLE 2. Quality and adequacy of the studied distributions in capturing the characteristics of the SRT variable as goodness-of-fit criteria

|     | Exponential | Gamma | Lognormal | Weibull | Loglogistic |
| --- | --- | --- | --- | --- | --- |
| AIC | 1508.387 | 1139.478 | 1106.023 | 1240.760 | 1074.187 |
| BIC | 1512.032 | 1146.769 | 1113.314 | 1248.051 | 1081.478 |

NOTE: AIC = Akaike's Information Criterion, BIC = Bayesian Information Criterion

According to the AIC and BIC results, the log-logistic distribution has the lowest values. This indicates that this distribution is the best-fitting theoretical distribution for the modelling of the SRT variable, and the lognormal shows the second smallest value among the fitted distributions, suggesting that it can also be considered a viable option for the modelling process.

### D. Model assumption checking and modelling

Before modelling the linear covariates with respect to SRT, it is important to assess the model and distribution assumptions to ensure that the chosen model and model's distributions are appropriate for the data and that the resulting estimates are reliable. To assess the appropriateness of the AFT model for the log-logistic and lognormal distributions, we created graphical illustrations (see Fig. 7). To examine the AFT log-logistic regression model we plotted $\log[\hat{S}(t)/(1-\hat{S}(t))]$ against $\ln(t)$ and observed that the levels of interaction types variable exhibit a pattern where they resemble nearly straight lines. The lines representing the different levels of interaction types show a partial parallelism, although one level deviates from the observed parallel pattern observed among the others.

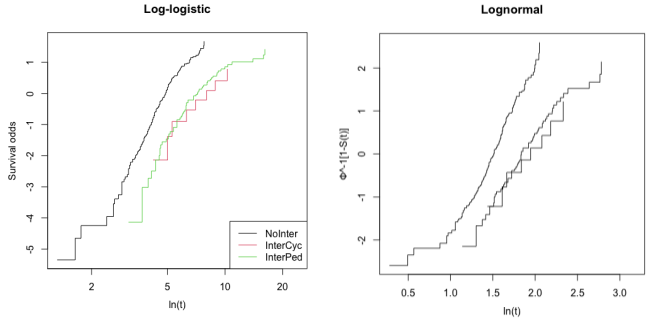

Figure 7. Graphical check of the AFT model assumption and both log-logistic and lognormal distributions for the modelling.

Hence, both the AFT model and the assumption of a log-logistic distribution are supported by the theoretical analysis. For the lognormal AFT regression model, in the graph, the linearity of $\Phi^{-1}[1 - \hat{S}(t)]$ plotted against $\ln(t)$ indicates that the AFT model with a lognormal distribution is suitable for the modeling task.

Based on the values presented in Table 3, the linear AFT model with log-logistic distribution emerges as the best model, as it shows the smallest AIC and BIC values.

TABLE 3. Goodness-of-fit criteria for the models

|  | AIC | BIC |
|---|---|---|
| **Lognormal** | 816.734 | 853.189 |
| **Log-logistic** | 731.930 | 768.384 |

NOTE: AIC = Akaike's Information Criterion, BIC = Bayesian Information Criterion

As evidenced by the model's p-value (almost 0), shown in Table 4, the AFT log-logistic model rejects the null hypothesis at a 0.05 significance rate. This indicates that there is a significant difference in the average speed reduction time among road users (i.e., cyclists and pedestrians) at the zebra crossing in an unsignalized intersection compared to the baseline scenario.

TABLE 4. The AFT log-logistic linear regression model's summary

|  | Coef | Std.error | Z stat | p-value |
|---|---|---|---|---|
| Intercept | 1.50079 | 0.10166 | 14.76 | < 2e-16 |
| $V_m$ | -0.12578 | 0.00677 | -18.58 | < 2e-16 |
| $LV_i$ | 0.01638 | 0.00221 | 7.40 | 1.3e-13 |
| $LV_m$ | 0.01568 | 0.00354 | 4.43 | 9.5e-06 |
| dav | -0.18975 | 0.02416 | -7.85 | 4.0e-15 |
| Mtype_turning_left | 0.22392 | 0.08731 | 2.56 | 0.01032 |
| Mtype_turning_right | 0.03948 | 0.02247 | 1.76 | 0.07890 |
| IntType_IntCyc | -0.00178 | 0.06854 | -0.03 | 0.97923 |
| IntType_IntPed | -0.15059 | 0.03961 | -3.80 | 0.00014 |
| Log(scale) | -2.41191 | 0.05312 | -45.41 | < 2e-16 |
| Scale = 0.0896 | | | | |
| Loglik(model) = -356 | | | | |
| Chisq =358.26 on 8 degrees of freedom p = 1.6e-72 | | | | |
| Number of Newton-Raphson Iterations: 6 | | | | |

The accelerated log-logistic failure time model's survival function is given as:

$$S(t) = \frac{1}{1+(\exp(\boldsymbol{\beta}'\mathbf{X})t)^p} \quad (7)$$

where p is a scale parameter. After solving the survival function for t and plugging the statistically significant coefficients, we obtain the following model on average time:

$$SRT = \exp(1.5 - 0.13V_m + 0.016LV_i + 0.015LV_m - 0.190\text{dav} + 0.22\text{left\_turning} - 0.15\text{IntPed})$$

Based on the model's analysis, an increase in $LV_i$ and $LV_m$ variables, as well as a left turn at an unsignalized intersection, results in an increase in the expected SRT. Conversely, an increase in $V_m$ and dav variables leads to a decrease in the expected SRT.

Based on the model's results, we observe that SRT(NoInter) can be approximated as exp(-0.15) times SRT(IntPed) or equivalently, 0.861 times SRT(IntPed). This implies that the average SRT of drivers in the absence of any interaction is estimated to be 4.48 seconds, whereas when drivers are likely to yield to pedestrians, the average SRT increases to 5.2 seconds. Furthermore, the estimated 95% confidence interval for the average SRT related to the yielding behavior towards pedestrians at the zebra crossing of an unsignalized intersection falls within the range of [4.82, 5.62] seconds.

The plotted empirical survival curve of SRT for each scenario by using the Kaplan-Meier estimator is illustrated in Fig. 8. As expected, the probability of encountering a longer speed reduction time while yielding to VRUs, and when drivers have no interaction with road users decreases as time elapses. However, as we can see from the plot, yielding behavior towards VRU tends to be more gradual and prolonged compared to the no-interaction scenario. For instance, after 6 seconds, there is approximately a 65% probability of experiencing a longer SRT for pedestrians, whereas in the absence of any interaction, the probability is less than 15%.

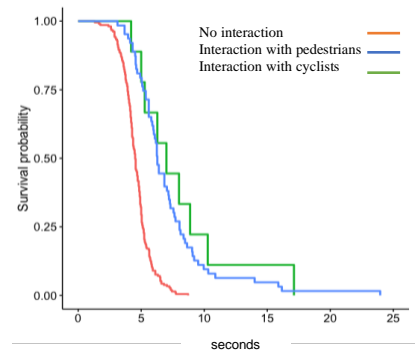

Figure 8. Survival function of drivers' yielding behavior towards vulnerable road users at unsignalized intersections, focusing on the duration in seconds The empirical survival function is estimated by the Kaplan-Meier method.

The effect of the naturalistic yielding behavior of drivers towards cyclists at the zebra crossing in the unsignalized intersection with respect to SRT is found to be statistically insignificant, as indicated by a p-value of 0.98. This suggests that there is no difference in the yielding behavior of drivers interacting with cyclists compared to drivers' no interaction and interaction with pedestrians scenarios. However, it is important to note that this result could potentially change with a larger sample size, which may provide more robust insights.

## V. Conclusion and future work

The visual analyses shown in this work support the notion that there are notable differences in the distribution of speed reduction times across different scenarios, emphasizing the need to account for these variations in the modelling process of autonomous vehicles. The further approach presented in this work allowed us to assess the statistical significance of the observed differences and gain insights into the unique characteristics of each yielding behavior scenario. Both the log-logistic and lognormal distributions exhibited favorable goodness-of-fit characteristics, making them suitable candidates for capturing the underlying patterns in the speed reduction time data.

## Acknowledgement

This work was supported by the Austrian Science Fund (FWF), project number P 34485-N. It was additionally supported by the Austrian Ministry for Climate Action, Environment, Energy, Mobility, Innovation, and Technology (BMK) Endowed Professorship for Sustainable Transport Logistics 4.0., IAV France S.A.S.U., IAV GmbH, Austrian Post AG and the UAS Technikum Wien

## References

[1] Schneider, R.J., Sanatizadeh, A., Shaon, M.R.R., He, Z. and Qin, X., 2018. Exploratory analysis of driver yielding at low-speed, uncontrolled crosswalks in Milwaukee, Wisconsin. Transportation research record, 2672(35), pp.21-32.

[2] Rothenbücher, D., Li, J., Sirkin, D., Mok, B., and Ju, W. "Ghost driver: A field study investigating the interaction between pedestrians and driverless vehicles," 2016 25th IEEE International Symposium on Robot and Human Interactive Communication (RO-MAN), New York, NY, USA, 2016, pp. 795-802.

[3] Lee, Y.K., Rhee, Y.E., Ryu, J.K. and Hahn, S., "Gentlemen on the Road: Understanding How Pedestrians Interpret Yielding Behavior of Autonomous Vehicles using Machine Learning." 2020.

[4] de Miguel, M. Á. Fuchshuber, d., Hussein, A. Olaverri-Monreal, C. "Perceived Pedestrian Safety: Public Interaction with Driverless Vehicles," 2019 IEEE Intelligent Vehicles Symposium (IV), Paris, France, 2019, pp. 90-95

[5] Alvarez, W. M., de Miguel, M.A., García, F. and Olaverri-Monreal, C. "Response of Vulnerable Road Users to Visual Information from Autonomous Vehicles in Shared Spaces," 2019 IEEE Intelligent Transportation Systems Conference (ITSC), Auckland, New Zealand, 2019, pp. 3714-3719

[6] Olaverri-Monreal, C., 2020. Promoting trust in self-driving vehicles. Nature Electronics, 3(6), pp.292-294.

[7] Alvarez, W.M., Moreno, F.M., Sipele, O., Smirnov, N. and Olaverri-Monreal, C., 2020, October. Autonomous driving: Framework for pedestrian intention estimation in a real world scenario. In 2020 IEEE Intelligent Vehicles Symposium (IV) (pp. 39-44).

[8] Bella, F. and Silvestri, M., "Survival model of drivers' speed reduction time at bicycle crossroads: a driving simulator study". Journal of Advanced Transportation, 2018.

[9] Sheykhfard, A., Haghighi, F., Papadimitriou, E. and Van Gelder, P., 2021. Analysis of the occurrence and severity of vehicle-pedestrian conflicts in marked and unmarked crosswalks through naturalistic driving study. Transportation research part F: traffic psychology and behaviour, 76, pp.178-192.

[10] Sun, S., Zhang, Z., Zhang, Z., Deng, P., Tian, K. and Wei, C., 2022. How Do Human-Driven Vehicles Avoid Pedestrians in Interactive Environments? A Naturalistic Driving Study. Sensors, 22(20), p.7860.

[11] Kalantari, A.H., Yang, Y., de Pedro, J.G., Lee, Y.M., Horrobin, A., Solernou, A., Holmes, C., Merat, N. and Markkula, G., 2023. Who goes first? a distributed simulator study of vehicle–pedestrian interaction. Accident Analysis & Prevention, 186, p.107050.

[12] Mohammadi, A., Dozza, M. and Bianchi-Piccinini, G., 2022. How Do Cyclists Negotiate an Unsignalized Intersection with a Vehicle? Modeling Cyclists' Yielding Behavior Using Naturalistic Data. Modeling Cyclists' Yielding Behavior Using Naturalistic Data (November 15, 2022).

[13] Bock, J., Krajewski, R., Moers, T., Runde, S., Vater, L. and Eckstein, L., 2020, October. The Ind dataset: A drone dataset of naturalistic road user trajectories at German intersections. In 2020 IEEE Intelligent Vehicles Symposium (IV) (pp. 1929-1934). IEEE.

[14] Kleinbaum, D.G. and Klein, M., 1996. Survival analysis a self-learning text. Springer.

[15] Moore, D.F., 2016. Applied survival analysis using R (Vol. 473). Cham: Springer.